\documentclass{pas}

\usepackage{multirow}
\usepackage{aas_macros}
\usepackage{xcolor}
\usepackage[cal=cm]{mathalpha}
\usepackage[dvipsnames]{xcolor}
\usepackage{dblfloatfix}
\usepackage{threeparttable}
\usepackage{hyperref}
\usepackage[caption=false]{subfig}
\usepackage{orcidlink}

\begin{document}

\lefttitle{Publications of the Astronomical Society of Australia}
\righttitle{Kotorashvili \& Blackman}

\jnlPage{1}{10}
\jnlDoiYr{2026}
\doival{10.1017/pasa.xxxx.xx}



\title{Observed x-ray spectral indices support  magnetic buoyancy sourced   coronae  in Seyfert active galactic nuclei}


\author{{Ketevan Kotorashvili}$^{1,2}$\orcidlink{0000-0002-5706-1299} and {Eric G. Blackman}$^{1,2}$\orcidlink{0000-0002-9405-8435}}
\affil{$^1$Department of Physics and Astronomy, University of Rochester, Rochester NY 14627 and
$^2$Laboratory for Laser Energetics, University of Rochester, Rochester, NY 14623, USA}

\corresp{K. Kotorashvili, Email: kkotoras@ur.rochester.edu}

\citeauth{Kotorashvili K and Blackman EG, Observed x-ray spectral indices support  magnetic buoyancy sourced   coronae  in Seyfert active galactic nuclei {\it Publications of the Astronomical Society of Australia} {\bf 00}, 1--10. https://doi.org/10.1017/pasa.xxxx.xx}

\history{(Received xx xx xxxx; revised xx xx xxxx; accepted xx xx xxxx)}

\begin{abstract}

Standard radiation spectra modeling of active galactic nuclei like Seyfert galaxies typically involves tuning the relative contributions from an optically thick disc, optically thin corona, dust torus, and outflow by hand to match observations.  The physics that quantifies this relative partitioning is not precisely understood. Here we make progress toward improving this understanding by testing the implications of a  conceptual paradigm that the corona is supplied from below only by magnetic fields whose scales are large enough to escape turbulent shredding on their buoyant rise time.  We use a generalized mean field accretion disc theory that includes both local radial angular momentum transport and angular momentum transport into a corona. We then use previous  numerical simulations of magnetized turbulent accretion disc energy spectra to connect the fraction of disc magnetic energy that would escape to the  fraction of coronal to total luminosity. 
Without explicitly calculating detailed particle acceleration and radiative processes, we  assume that the disc radiation spectrum is thermal and the coronal radiation spectrum is a power law. We then   show that the aforementioned  paradigm for corona formation constrains the allowed x-ray power law index to a narrow range which agrees with observations. For the three AGNs studied here the predicted mean photon index is $\overline{\Gamma}=1.76$, consistent with the mean observed value of $\overline{\Gamma}_{\rm obs}=1.73$, bolstering confidence in the paradigm.

\end{abstract}


\maketitle

\section{Introduction}

The observed spectra of Seyfert active galactic nuclei (AGN) have long shown that accretion into a black hole cannot be described exclusively by thermal UV disc emission, but that 20-40\% of their emission comes in x-rays. This high-energy component is generally associated with a hot, optically thin corona, whose structure and physical origin are still not fully understood.

The thermal component of the spectrum is usually described using the geometrically thin, optically thick accretion disc model of \cite{SS1973} (SS73). In this framework, angular momentum is transported radially through turbulent stress, and energy is dissipated locally within the disc and radiated from its surface.  The magentorotational inability (MRI) is widely thought to be a  plausible source for MHD turbulence in the accretion disc and for the radial transport of angular momentum in this framework, but other transport mechanisms including global MHD instabilities \citep{Fatima+Matthew2022} and large scale magnetic torques  \citep{Blandford+Payne1982,Jacquemin-Ide2021} may also contribute substantially to the transport. Recently, \cite{Vyas2026} showed that luminous, compact corona can generate a radiation-driven poloidal field which can be further amplified by differential rotation providing an additional magnetization mechanism that can be faster than MRI. Large scale magnetic fields, in particular, transport angular momentum and energy vertically away from the disc to form the corona and wind or jet. \cite{Blackman+Nauman2015} discussed the importance of including these large-scale and vertical transport processes in accretion disc modeling in order to self-consistently describe the disc, corona and outflows and to predict their relative contributions to the observed AGN spectra. 

An important step toward extending standard mean-field disc theory was made by \cite{Kuncic+Bicknell2004, Kuncic+Bicknell2007}. They relaxed some of the assumptions of the standard $\alpha$-disc model (SS73) by considering both radial transport due to local stresses and vertical transport of angular momentum associated with large scale MHD stresses. This allows part of the accretion power to be removed from the disc instead of being dissipated therein.  While their framework accounts for the energy removed from the disc via outflows, lowering the total disc emission,  it does not address what fraction  the accretion power is dissipated in the corona or predict the resulting x-ray spectrum.

Understanding how vertically transported energy contributes to the x-ray emission requires considering both physical processes responsible for forming the corona and its geometry.  A plausible formation scenario is that magnetic fields generated in the accretion disc  rise buoyantly above the disc, carrying a fraction of accretion energy and forming coronal loops \citep{Galeev+1979, Field+Rogers1993}. This magnetic energy can then be released through magnetic reconnection, which provides an acceleration environment that may include shocks, stochastic, or direct electric field  acceleration of  electrons \citep{Blackman+Field1994}. The accelerated electrons then inverse Compton scatter thermal photons from the disc to produce  the observed x-ray  emission \citep{Haardt+Maraschi1991,Field+Rogers1993}. \cite{Blackman+Pessah2009} showed that if the corona is powered by buoyantly rising magnetic structures generated within the turbulent disc, then those structures should be large enough to survive buoyant rise without being shredded by the disc turbulence. The competition between buoyant rise and turbulent shredding therefore sets a critical scale for magnetic structures that can reach the corona. They related the observed ratio of coronal to bolometric emission to the fraction of magnetic energy that rises to the corona, inferring  that a substantial fraction of magnetic energy in accretion discs is produced in large-scale fields.

The coronal geometry is still an active area of study, and several different geometries have been proposed. In compact geometries such as lamppost models, corona is concentrated in a compact region close to the spin axis of the black hole, while in slab or sandwich geometries the corona extends  above and below much or all of the full accretion disc \citep{Wilkins+Fabian2012, Poutanen2018}. Here the corona can originate from magnetic instabilities (e.g. \cite{DiMatteo1998}). In spherical or quasi-spherical configurations the corona surrounds the inner disc with the disc truncated further out. Other proposed geometries include conical, funnel-shaped, and cylindrical corona geometries, which are usually associated with the base of a jet  (e.g. \cite{Henri+Pelletier1991,Henri+Petrucci1997,Markoff+Nowak2005} or a failed jet \citep{Ghisellini+Haardt2004}, and a patchy coronal model consisting of localized coronal regions above the disc \citep{Kuncic+blackman+fabian1996,Kim+DiGesu2024,Meng+You2025}.  

Different coronal geometries produce different observational signatures. The size of the corona and location relative to the disc determines how much of the disc radiation is Comptonised, the shape of the Comptonised continuum, the strength and shape of the reflected emission \citep{Wilkins+Fabian2012}, and the time-lags between direct and reflected photons \citep{Wilkins2016}. The angle of polarization of the x-ray emission also depend on the coronal geometry \cite{Gianolli2023}. These observables can all be used to constrain the geometry of the corona \citep{Wilkins+Fabian2012, Wilkins2016,Ursini2022}.

In this work we build on the framework \cite{Kuncic+Bicknell2007} by connecting the vertically transported magnetic energy to the magnetic buoyancy picture of \cite{Blackman+Pessah2009} and then to the coronal emission and spectra.  We use magnetic energy spectra from MHD simulations \citep{Held+Mamatsashvili+Pessah2024} to quantify the fraction of magnetic energy contained in large scale structures that survive the buoyant rise and determine how much energy is removed from the disc and transported vertically. This allows us to predict the spectral index for the coronal power law spectrum.
In our framework, the coronal energy is supplied by  magnetic structures rising from the disc, and we assume a geometrically thin, disc- or slab-like corona that  extends from  the innermost stable circular orbit (ISCO) to an outer radius.

The paper is organized as follows. In Section \ref{sec:model intro}, we introduce the model framework, including radial and vertical transport of angular momentum \citep{SS1973,Kuncic+Bicknell2004,Kuncic+Bicknell2007,Blackman+Nauman2015}, the escape condition for magnetic structures \citep{Blackman+Pessah2009}, the resulting disc flux. In Subsection \ref{sec:our model}, we generalize disc flux to include the corona and present our disc-corona model. In Section \ref{sec:spectra intro}, we discuss spectral implications of the model. We first characterize the Seyfert spectra used in this paper and then use observed x-ray data to constrain the model parameter space and test the consistency of the model framework with observations. In Subsection \ref{sec:single data point}, we 
demonstrate a practical use of the model to recover the coronal spectrum from only a single x-ray data point. In Subsection \ref{sec:microphyscis} we discuss how predicted photon index compares  with results obtained from microphysical models of particle acceleration and recent simulations and conclude in Section \ref{sec:conclusion}.

\section{Model framework }\label{sec:model intro}
\subsection{Inclusion of radial and vertical transport}
The standard  axisymmetric  Shakura-Sunyaev accretion theory that invokes turbulent viscous transport \citep{SS1973} of angular momentum  emerges from a formal density-weighted, vertically and azimuthally averaged  mean-field theory.  It  approximates the radial angular momentum  transport in an accretion disc \citep{Balbus+1998} but  does not  account for large-scale or  vertical transport of angular momentum \citep{Blackman+Nauman2015}. 
\cite{Kuncic+Bicknell2004,Kuncic+Bicknell2007} provide a mean-field extension that couples the radial and vertical transport of mass, momentum, and energy. We adapt and augment this approach here.

Physical quantities are decomposed into mean and fluctuating components using density-weighted statistical averaging over azimuth, leaving dependencies on radius and height. Vertical integration produces mean quantities that depend only on radius.  Thus, for a quantity $a(r,\phi,z)$, this means 
\begin{equation}
{A}(r) = \int_{-h}^{+h} \langle a(r,z) \rangle dz=\frac{\int_0^{2\pi}\int_{-h}^{+h}\rho(r,z,\phi) a(r,\phi,z) dz d\phi}{2\pi\Sigma(r)},
\label{1}
\end{equation}
where $\rho$ is the density, $\Sigma=\int_{-h}^{+h}\rho dz$ is the surface density, where $h$ is disc scale height and with $z=\pm h$,
and capital $A$ indicates the mean quantity dependent only on radius. Hereafter $\langle a(r,z) \rangle$ is used to indicate the density-weighted azimuthal average of quantity $a$.
In this formalism, vertical transport can remove angular momentum 
and energy via magnetic stresses and/or outflows 
which can in turn modify the 
thermal disc emission spectra of standard optically thick,  radial-only transport models,  where  energy dissipation occurs only in the disc.

However, \cite{Kuncic+Bicknell2004} and \cite{Kuncic+Bicknell2007} consider that the energy associated with vertical transport is simply removed from the emission spectrum and depletes the thermal disc emission. Here we instead use that 
energy as source for  coronal emission, and thus the 
coronal emission spectrum.
Our approach is anchored by the paradigmatic ansatz that the corona is supplied from below only by magnetic fields whose scales are large enough to escape turbulent shredding on their buoyant rise time \citep{Blackman+Pessah2009}.
These large-scale magnetic fields dissipate  in the corona
and power the observed x-ray emission.  The detailed physics of this dissipation, electron acceleration and heating and inverse Compton have long been topics of previous  and active research. For present purposes however, 
we need only recognize  that there is a coronal contribution to the spectrum and a thermal disc contribution, and our work here addresses constraints on their relative contributions.

\subsection{Escape condition for magnetic structures}\label{sec:f_s and q}
\cite{Blackman+Pessah2009} show how to quantify the observed ratio of coronal to bolometric emission to the fraction of magnetic energy produced in the disc that is large enough to escape to the corona. For magnetic fields to reach the corona before dissipating   
the buoyant rise  time $t_{\rm b}$ must be shorter than the turbulent diffusion timescale $t_{\rm d}$. These timescales can be estimated, respectively, as
\begin{equation}
    t_b \equiv \frac{H}{U_b}\label{eq:t_b}
\end{equation}
where $H$ is the disc half thickness and  $U_{\rm b}$ is the characteristic buoyancy speed and 
\begin{equation}
    t_d \equiv \frac{l^2}{\nu_t}\label{eq:t_d}
\end{equation}
where $l$ is the length scale and $\nu_{\rm t}$ is turbulent diffusion coefficient. The latter can be crudely approximated as  $\nu_{\rm t} \sim \upsilon l_{\rm t}$ where $\upsilon$ is the dominant turbulent speed and $l_{\rm t}$ is the characteristic energy containing turbulent scale. In the $\alpha$-viscosity thin disc framework (SS73) this can be further approximated as $ \upsilon l_{\rm t} \simeq \alpha c_{\rm s} H$, where $c_{\rm s}$ is a sound speed. From the escape condition  $t_{\rm b}< t_{\rm d}$, the critical  scale that magnetic structures must have in the disc to escape to the corona or outflows is 
\begin{equation}
l^2>l^2_c\equiv \alpha H^2\frac{c_s}{U_b}.\label{eq:lc}
\end{equation}
The fraction of disc magnetic energy $f_{\rm s}$ with scales larger than $l_{\rm c}$ can be written in terms of the integrated disc magnetic energy spectrum
\begin{equation}
f_s =
\frac{
\int_{k_c}^{k_{\max}} E_M(k)dk
}{
\int_{k_{\min}}^{k_{\max}} E_M(k)dk
},\label{eq:fs}
\end{equation}
where $E_M(k)$ is the magnetic energy spectrum, the integration bounds  are $k_{\rm c} =2 \pi/l_{\rm c}$, $k_{\rm min} =2 \pi/H$ and $k_{\rm max} =2 \pi/l_{\rm diss}$, and $l_{\rm diss}$ is the dissipation length scale. 

From \cite{Blackman+Pessah2009}, the ratio $q$ of the energy dissipation flux of large-scale fields in the corona $F_c$ to total dissipation flux  is given by $q=F_c/(F_d+F_{\rm up})$, where  $F_d$ is the energy flux associated with the dissipation of small-scale (sub-critical scale) fields that dissipate in the disc and $F_{\rm up}=F_c+F_w$ energy dissipation associated in the corona or with the outflows.
The ratio $q$ is also interpreted as an estimate of the observed ratio of coronal to bolometric emission, 
The disc energy dissipation flux is given by
\begin{equation}
F_d \equiv Q \Sigma \Omega^2 \nu_t (1 - f_s)
= 2 Q \alpha \beta c_s \epsilon_{\rm mag} (1 - f_s),
\label{eq: F_d}
\end{equation}
where $\Sigma$ is the surface density; $\Omega$ is the local angular speed; $\beta\equiv \rho c_s^2/2\epsilon_{\rm mag}$, where $\rho$ is average disc mass density and $\epsilon_{\rm mag}\equiv B^2/8\pi$ is the disc magnetic energy density; and $Q= (d {\rm ln} \Omega/dR)^2R^2$. We have  also corrected a typo in the definition of $Q$ given in \cite{Blackman+Pessah2009}.
The energy dissipation flux above disc  is given by
\begin{equation}
F_{\rm up} \equiv f_s U_b \epsilon_{\rm mag}.
\label{eq: F_up}
\end{equation}

Using equations (\ref{eq: F_d}) and (\ref{eq: F_up}), the relation between $q$ and  the fraction of magnetic energy produced in the disc that is large enough to escape from the disc is then
\begin{equation}
 \frac{q}{1 - q}
= \frac{f_s}{2 Q \alpha \beta (1 - f_s)} \frac{U_b}{c_s}.\label{eq:qandfs}
\end{equation}
In the present framework, we use this relation by first determining $f_s$ from  numerical simulations of magnetized turbulent accretion discs and then constrain the corresponding fraction of coronal to total luminosity and the coronal radiation spectrum. From equation (\ref{eq:lc}) for both the maximum buoyancy velocity $U_b \sim  c_s$  and the minimum  buoyancy velocity $U_b \sim \alpha c_s$ \cite{}Vishniac (1995)  with  $\alpha < 1$ we have $l_c \leq H$ implying that  $f_s \leq 1$. Generally  $q 
\leq f_s \leq 1$ , but since some of the magnetic energy can be be lost into outflows rather than  radiated directly by the corona, $q 
<f_s \leq 1$.

\subsection{Disc flux}\label{sec:kuncic disc}
Following the \cite{Kuncic+Bicknell2004,Kuncic+Bicknell2007} treatment of coupled radial and vertical angular momentum
transport, we adopt a mean-field MHD description of a geometrically thin, non-self gravitating disc. We give
the expression for the disc radiative flux here as the starting point for incorporating our new ingredients. The disc equations are vertically integrated over the disc thickness.  Angular momentum can be transported radially by local viscous stresses, 
and vertically from both mass loss by a wind and/or magnetic stresses on the disc surface. 
This approach leads to the mean-field radiative flux of the disc given by 
\begin{align}
F_{\rm d}(r) & \nonumber \approx
\frac{3GM\dot{M}_a(r)}{8\pi r^3}
\left[
1 - \frac{\dot{M}_a(r_i)}{\dot{M}_a(r)}
\left(\frac{r_i}{r}\right)^{1/2}
\right] \\ \nonumber
&\quad - \frac{3}{4}
\left(\frac{r_i}{r}\right)^2 T_{r\phi}(r_i)\Omega \\ 
&\quad - \frac{3\Omega}{8\pi r^2}
\int_{r_i}^{r}
\left[
r^2 \Omega \frac{d\dot{M}_a}{dr}
- 4\pi r^2 \langle t_{\phi z} \rangle
\right] dr,\label{eq:disc flux_kuncic}
\end{align}
where $G$ is the gravitational constant, $M$ is the black hole mass and $\dot{M}_a(r)$ is the mass accretion rate. The first term on the right resembles energy dissipation from the standard disc theory, but with a  radially varying accretion rate due to vertical mass loss
that can exceed the value at the inner most stable orbit $r_i$ such that $\dot{M}_a(r)\ge \dot{M}_a(r_i)$.
 The second term on the right describes the rate at which nonzero vertically averaged MHD stresses $T_{r\phi}$ dissipate turbulent energy at  $r_i$, with
 {$T_{r\phi}=\int_{-h}^{+h}\langle t_{r \phi}\rangle dz$.}, where $\langle t_{r \phi}\rangle$ is the density-weighted azimuthal average of the combined magnetic and Reynolds stresses.
 This vanishes for a zero stress inner boundary. 
 The final term  of equation (\ref{eq:disc flux_kuncic})
 describes the rate at which energy is removed from the disc due to  mass outflow and MHD stresses $\langle t_{\phi z} \rangle$, thereby including both magnetic and Reynolds stresses.
\begin{table*}[!b]
\centering
\begin{threeparttable}
\caption{Physical parameters adopted for ARK120, MCG63015 and NGC4593:} 
\label{tab:agn}
\begin{tabular}{lcccc}
\toprule
Parameter & ARK120\textsuperscript{\hyperlink{noteA}{a}} & MCG63015\textsuperscript{\hyperlink{noteB}{b}} & NGC4593\textsuperscript{\hyperlink{noteC}{c}} & Determined by\\
\midrule
\hline
$M_{\rm BH}$ ($M_\odot$)  &
$1.5 \times 10^{8}$ &
$4.5 \times 10^{6}$  &
$8.0 \times 10^{6}$ & Observationally Inferred\\
$L_{\rm bol}$ ($\rm{erg\,s^{-1}}$)  &
$2 \times 10^{45}$ &
$2.0 \times 10^{44}$ &
$1.0 \times 10^{44}$ & Observationally Inferred\\
$a$ &
$0.83$ &
$0.98$ &
$0.99$ & Spectral modeling\\
$\dot{M}$ ($M_\odot\,\rm{yr^{-1}}$) &
$0.272$ &
$0.015$ &
$0.0067$ & Theoretically estimated form $L_{\rm bol}$\\
$p$ &
$0.11$ &
$0.08$  & 
$0.07$ & Theoretically estimated form $L_{\rm bol}$ \textsuperscript{\hyperlink{noteD}{d}}  \\
$r_{\rm i}$ ($r_g$) &
$2.75$ &
$1.61$ &
$1.45$ & Theoretically estimated form $a$\\
$r_{\rm out}$ ($r_g$) &
$10^{3}$ &
$10^{3}$ &
$10^{3}$ & Model input \\
$E_{\rm cut} (\rm kev)$ &
$7 \times 20^{19}$ &
$7 \times 20^{19}$  &
$7 \times 20^{19}$ & Model input \\
\hline
\bottomrule
\end{tabular}

\vspace{0.3cm}
\begin{tablenotes}[flushleft]
\footnotesize
\item[a] \hypertarget{noteA}{}%
References for $M_{\rm BH}$, $L_{\rm bol}$ and $a$ \citep{Porquet2019}.
\item[b] \hypertarget{noteB}{}%
References for $M_{\rm BH}$, $L_{\rm bol}$ \citep{McHardy2005} and $a$ \citep{Brenneman+Reynolds2006}.
\item[c] \hypertarget{noteC}{}%
References for $M_{\rm BH}$, $L_{\rm bol}$ \citep{Ruschel-Dutra2021}; $a$ is unconstrained, but we follow  \citep{Xiang2022}.
\item[d] \hypertarget{noteD}{}%
Detailed calculations can be found in Appendix \ref{app:p}.
\end{tablenotes}
\end{threeparttable}
\end{table*}
\subsection{Generalization of flux for disc with a corona}\label{sec:our model}

Here we summarize the main new ingredients of our framework which differ from that of \cite{Kuncic+Bicknell2007}.
Unlike in their model,  we do not introduce a critical radius beyond which mass loss vanishes.  Instead, we adopt a  mass loss prescription starting at the outer boundary
\begin{equation}
\dot{M}_a(r)=\dot{M}\left(\frac{r}{r_{\rm out}}\right)^p,\label{eq:mass loss}
\end{equation}
where $\dot{M}$ is a net accretion rate at $r_{\rm out}$, $0<p<1$ is the mass loss index  \citep{Blandford+Begelman1999,Kuncic+Bicknell2004,Kuncic+Bicknell2007}.
Also, we do not prescribe the relative importance of outflows in transporting angular momentum to decrease with increasing radius through an additional radial power-law dependence as in \cite{Kuncic+Bicknell2004}. Here, we instead adopt a physical prescription in which a constant fraction  $f_s$ of the magnetic energy is extracted vertically from the disc to the corona and outflows at each radius.
Under these assumptions, the  mean-field radiative flux from the disc is reduced by the factor ($1-f_s$), leading to:
\begin{align}
F_{\rm d}(r) =
\frac{3GM\dot{M}_a(r)}{8\pi r^3}
\left[
1 - \frac{\dot{M}_a(r_i)}{\dot{M}_a(r)}
\left(\frac{r_i}{r}\right)^{1/2}
\right] (1-f_{\rm s}).\label{eq:disc flux}
\end{align}
Plugging the mass loss prescription given by equation (\ref{eq:mass loss}) into equation (\ref{eq:disc flux}), we get: 
\begin{align}
F_{\rm d}(r) =
\frac{3GM\dot{M}}{8\pi r^3}
\left[
1 - \left(\frac{r_i}{r}\right)^{p+1/2}
\right] (1-f_{\rm s}).\label{eq:disc flux}
\end{align}
Integrating this flux over   disc area gives the total disc luminosity 
\begin{align}
L_d &= 2\int_{r_i}^{\infty} F_d(r) 2\pi r dr \\ \nonumber
&=\frac{1}{2}\frac{GM\dot{M}}{r_i}\left(\frac{r_{\rm out}}{r_i}\right)^{-p}\left(\frac{1+2p}{1-p}\right)(1-f_{\rm s}).\label{eq:total disc luminosity}
\end{align}
To compare our solutions with those of \cite{Kuncic+Bicknell2007}, we rewrite our solution in the form:
\begin{equation}
L_d = P_a - P_{\rm up},
\end{equation}
where $P_a$ is total accretion power in the disc and $P_{\rm up}$  total power removed from the disc and transported vertically into the corona and outflows. In our model, the total accretion power $P_a$ has a simpler expression and is given by
\begin{equation}
  P_a= \frac{1}{2}\frac{GM\dot{M}}{r_i}\left(\frac{r_{\rm out}}{r_i}\right)^{-p}\left(\frac{1+2p}{1-p}\right),\label{eq:power total}
\end{equation}
while the total power transported into the corona and outflows is:
\begin{equation}
    P_{\rm up}=f_s\frac{1}{2}\frac{GM\dot{M}}{r_i}\left(\frac{r_{\rm out}}{r_i}\right)^{-p}\left(\frac{1+2p}{1-p}\right)=f_s P_a. \label{eq:power corona}
\end{equation}
But we need to account for the fact that not all of the power removed from the disc is necessarily radiated by the corona. Part of this energy may be lost through additional sinks, such as outflows.  Following the  interpretation of the parameter $q$ discussed in \cite{Blackman+Pessah2009} and section \ref{sec:f_s and q}  the coronal luminosity can therefore written as:
\begin{equation}
    L_c=qP_a=\frac{q}{1-q}L_d\label{eq:L_c}
\end{equation}
In the limiting case where $2 Q \alpha \beta  U_b/c_s=1$, we recover $q=f_s$, corresponding  to a system without additional outflow sinks above the corona. But as mentioned in the previous section, realistic systems may have some fraction of the magnetic energy  carried away and dissipated in outflows rather than contributing directly to the coronal luminosity.

\section{Spectral Implications}\label{sec:spectra intro}
\subsection{Characterization of Seyfert spectra}\label{sec:seyfert spectra}

For  spectral calculations, we assume that the disc radiates locally as blackbody and that  the corona is responsible for  x-ray emission. The total disc radiation spectra is calculated by summing up the local blackbody emission $B_\nu$ from each local annulus:
\begin{equation}
L_\nu = 2\int_{r_i}^{\infty} \pi B_\nu[T(r)] 2\pi r dr
\label{17}
\end{equation}
The effective disc temperature $T(r)$  of each annulus is determined from the mean-field radiative flux through Stefan-Boltzmann relation:
\begin{equation}
T(r)= (F _d(r)/\sigma)^{1/4},
\label{18}
\end{equation}
where $\sigma$ is the Stefan-Boltzmann
constant.
Using equation (\ref{18}) to replace temperature by radial flux in (\ref{17}), the total disc spectrum can be written as:
\begin{equation}
L_\nu = 2\int_{r_i}^{\infty} \pi B_\nu\left[\frac{F _d(r)}{\sigma}\right]^{1/4}2\pi r dr
\end{equation}

In detailed microphysical models \cite{Field+Rogers1993,Haardt+Maraschi1991} inverse Compton scattering of soft accretion disc photons by coronal electrons thought to be the dominant process in the corona and responsible for producing observed hard x-ray power law emission. AGN x-ray spectra are generally considered as a mixture of thermal and non-thermal electron populations but the relative contribution of these components depends on which particle acceleration and energy dissipation processes operate in the corona,  but the relative fraction of thermal versus non-thermal electron components and what determines this fraction remains an active area of research \citep{Laha2025}. Some estimates suggest 30$\%$ of non-thermal particles \cite{Fabian2017}.

Regardless of the underlying microphysics, AGNs and x-ray binaries have x-ray spectra well described by a power law with a high energy cutoff. Here we simply adopt this phenomenological power-law approximation to describe coronal x-ray emission \citep{Rybicki+Lightman1979}. The corresponding photon number spectrum is given by:
\begin{equation}
 N_\nu=K\nu^{-\Gamma}*e^{-\nu/\nu_{\rm c}}, \label{eq:N_nu}
\end{equation}
where $K =\nu_{*}^{-\Gamma}  N_{\nu,0} $, $N_{\nu,0}$ is normalization of the power law  determined from the observed luminosity and $\nu_*$ is a normalization frequency, $\Gamma$ is photon index, $\nu_{\rm c}=E_c/h$ where $E_c$ coronal high energy cutoff and $\nu_{\rm c}$ is a corresponding frequency.

\subsection{Constraining model self-consistency with data} \label{sec:constraining model}
We focus on Seyfert 1 galaxies because their spectra are directly observable and their discs are amenable to thermal pressure rather than radiation dominated discs. We use observed x-ray data for three different Seyfert 1 AGNs (ARK120, MCG63015, and NCG4593) with different black hole masses, bolometric luminosities, spins and, mass accretion rates  to test our model  by constraining $\Gamma$. 
The specific AGN properties are summarized in Table \ref{tab:agn}. These quantities determine the total gravitational binding energy released through accretion. But to calculate the radius-dependent radiative flux of the disc, we need to take  energy fraction $f_s$ that is transported into the corona and/or outflows rather than dissipated in the disc.
To estimate $f_s$, we appeal to  magnetic energy spectra from numerical simulations \cite{Held+Mamatsashvili+Pessah2024}, and calculate $f_s$ using equation (\ref{eq:fs}) for the minimum and maximum buoyant velocity limits $U_{b, \rm min}=\alpha c_s$, $U_{b, \rm max}=c_s$ for the ranges of $\alpha$, $\alpha \beta$ which are given in Table \ref{tab:parameter_ranges}, along with ranges of $f_s$ and $q$ estimated from equation (\ref{eq:qandfs}). We keep only physical solutions $f_s>q$ and list their total number as $n$. More details on the calculations determining $f_s$  are given in Appendix \ref{app:f_s}.

\begin{table}[!t]
\begin{threeparttable}
 \caption{Quantities $\alpha$, $\beta$, $\alpha\beta$, $f_s$, $q$, $\Gamma$ and $n$, where $n$ represents number of physical solutions, for the full range and constrained ranges that result from   physically constraining $\alpha\beta$ and $\alpha$ as described in the text. We also list the dominant ranges of $f_s$, and  $q$, and the mean values of photon index $\overline{\Gamma}$ and photon-spectrum normalization $ \overline{N}_{\nu,0}$ for each individual AGNs and for the combined solutions calculated from all three. The normalization is evaluated at $\nu_*=10^{17} \rm Hz$ and has units of $\rm photons\, s^{-1}\, Hz^{-1}$. For the combined solutions we also list  $\overline{\Gamma}$ and $ \overline{N}_{\nu,0}$ intervals extended half of a standard deviation.}\label{tab:parameter_ranges}
\begin{tabular}{@{}l@{\hspace{1pt}}c@{\hspace{3pt}}c@{}}
\hline
parameter &
full range &
constrained range \\
\hline
$\alpha$        & $0.01 - 0.5$      & $0.06 - 0.21$ \\
$\alpha\beta$   & $0.1 - 1.0$       & $0.45 - 0.55$ \\
$\beta$         & $0.45 - 100.00$   & $2.14 - 9.17$ \\
$f_s$           & $0.140 - 0.920$   & $0.140 - 0.620$ \\
$q$             & $0.006 - 0.919$   & $0.008 - 0.617$ \\
$\Gamma$        & $0.010 - 4.306$   & $0.011 - 4.109$ \\
$n$       & $81,699$          & $10,536$ \\
\hline
dominant $f_s$\textsuperscript{\hyperlink{noteA}{a}}
 & $0.140 - 0.920$ & $0.360 - 0.620$ \\
dominant $q$\textsuperscript{\hyperlink{noteA}{a}}
 & $0.058 - 0.857$ & $0.322 - 0.592$ \\
\hline
ARK120 $\overline{\Gamma},\, \overline{N}_{\nu,0}$& $1.682,\,2.279\times10^{35}$
 & $1.728,\,3.447\times10^{35}$ \\
MCG63015 $\overline{\Gamma},\,\overline{N}_{\nu,0}$& $1.726,\,1.784\times10^{34}$
 & $1.864,\,4.135\times10^{34}$ \\
NGC4593 $\overline{\Gamma},\,\overline{N}_{\nu,0}$& $1.801,\,9.094\times10^{33}$
 & $1.673,\,1.070\times10^{34}$ \\
\hline
\multicolumn{3}{l}{All three AGN combined} \\
\hline
$\overline{\Gamma},\ \overline{N}_{\nu,0}$& $1.736,\quad 2.579\times10^{34}$
 & $1.763,\quad 4.642\times10^{34}$ \\
$\overline{\Gamma}\pm\sigma/2$
 & $1.36\leq\Gamma\leq2.10$
 & $1.38\leq\Gamma\leq2.14$ \\
$\overline{N}_{\nu,0}\pm\sigma/2$& $9.9\times10^{33}\leq N_{\nu,0}\leq1.1\times10^{35}$& $1.5\times10^{34}\leq N_{\nu,0}\leq1.8\times10^{35}$\\
\hline
\end{tabular}

\vspace{0.3cm}
\begin{tablenotes}[flushleft]
\footnotesize
\item[a] \hypertarget{noteA}{}%
Dominant ranges correspond to the central 95\% percentile
of the retained solutions.   
\end{tablenotes}
\end{threeparttable}
\end{table}
Using values from Table \ref{tab:parameter_ranges},  we calculate the total disc radiation spectrum and corresponding luminosity as discussed in section \ref{sec:our model}. Then, from equation (\ref{eq:L_c}), knowing both $q$ and the disc luminosity, we can find the corresponding coronal luminosity for each set of parameters. We then predict the coronal x-ray spectra and the corresponding photon index $\Gamma$.  
We use each observed x-day data point\footnote{Data obtained from the NASA/IPAC Extragalactic Database (NED)  \url{https://ned.ipac.caltech.edu}} independently to  normalize the photon energy distribution and obtain a set of power-law curves that go through the point.
Also the integrated luminosity under each acceptable power-law solution  matches the calculated coronal luminosity from equation (\ref{eq:L_c}), which then constrains the allowed values for the photon index $\Gamma$. This allows us to determine the corresponding coronal spectra power law, and in turn predict the disc-corona spectral energy distribution for each source.

\begin{figure}
    \centering
    \includegraphics[width=\columnwidth]{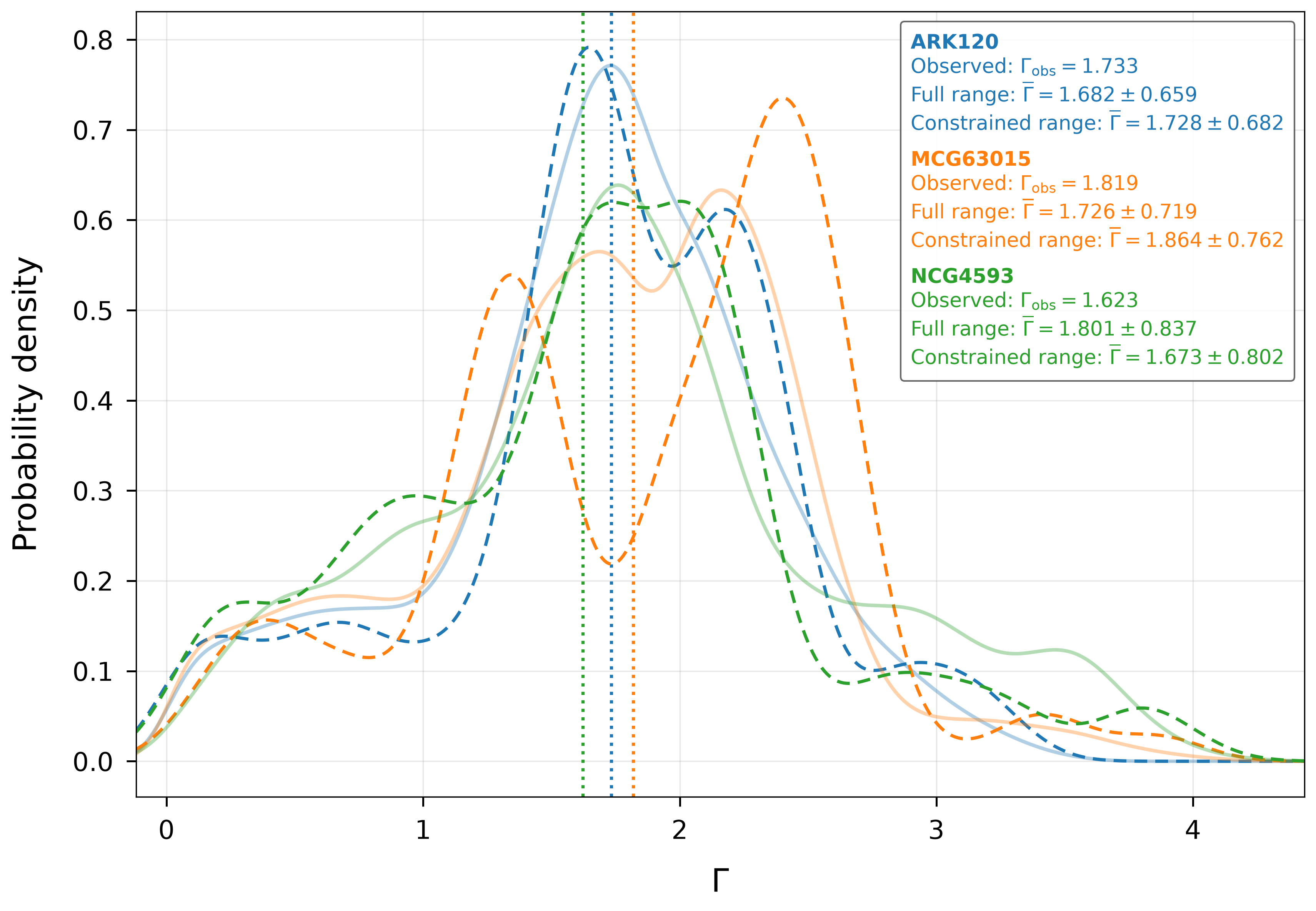}
    \caption{Probability distribution of the photon index $\Gamma$ for ARK120 - blue , MCG63015 - orange, and NCG4593 - green. Solid curves show the distribution for all physical solutions corresponding to the full parameter ranges in Table \ref{tab:parameter_ranges}, while dashed curves show distributions for the constrained parameter ranges. The vertical dotted lines delineate the best-fit observed photon indices $\Gamma_{\rm obs}$. The legend summarizes the values for $\Gamma_{\rm obs}$, and the mean values of $\Gamma$ and $N_{\nu,0}$ before and after applying the  parameter range constraints for each AGN.}
    \label{fig:Gamma distribution peragn}
\end{figure}

We apply this procedure to all available x-ray data points for ARK120, MCG63015, and NCG4593 AGNs for the full $\alpha$, $\alpha \beta$ parameter range given in Table \ref{tab:parameter_ranges}. Figure \ref{fig:Gamma distribution peragn} shows the $\Gamma$ distribution for each AGN individually and the corresponding mean values for the photon index $\overline{\Gamma}$. For each set of parameters and available x-ray data points, we can find zero, one or two possible values of $\Gamma$. We keep all valid solutions in our analysis. For some cases, no values of $\Gamma$ can simultaneously satisfy the normalization to the observed x-ray data point and the calculated coronal luminosity.  Appendix \ref{app:gamma} gives a more detailed description of how to find $\Gamma$, including the equation used to determine power law solutions and why multiple solutions can exist.

We compare the predicted statistically determined $\overline{\Gamma}$ with values independently inferred directly from the observed x-ray spectra $\Gamma_{\rm obs}$. For each AGN, we determine $\Gamma_{\rm obs}$ by fitting a power law to the observed x-ray data.  This procedure   allows us to (i) test the validity of a theoretically restricted  $\alpha \beta$ parameter range  and 
(ii) constrain the range of $\alpha$ itself 
by determining whether these restrictions improve the match of $\overline{\Gamma}$ to $\Gamma_{\rm obs}$. 
We first use the $\alpha\beta$ restriction  from 
\cite{Blackman+2008}, who find theoretically that $\alpha\beta$  should be approximately $0.5$, which motivates a restricted range $0.45 \leq \alpha\beta \leq 0.55$. 
We then  constrain $\alpha$ to $0.06 \leq \alpha\leq 0.21$ by requiring  the predicted mean photon index $\overline{\Gamma}$  to be consistent with the observed value $\Gamma_{\rm obs}$. Table \ref{tab:parameter_ranges} shows how these constraints change the number of solutions and the ranges of  $f_s$ $q$ and $\Gamma$ together with the dominant ranges of $f_s$ and $q$ and the $\overline{\Gamma}\pm\sigma/2$ and $\overline{N}_{\nu,0}\pm\sigma/2$ intervals.
\begin{figure}
    \centering
    \includegraphics[width=\columnwidth]{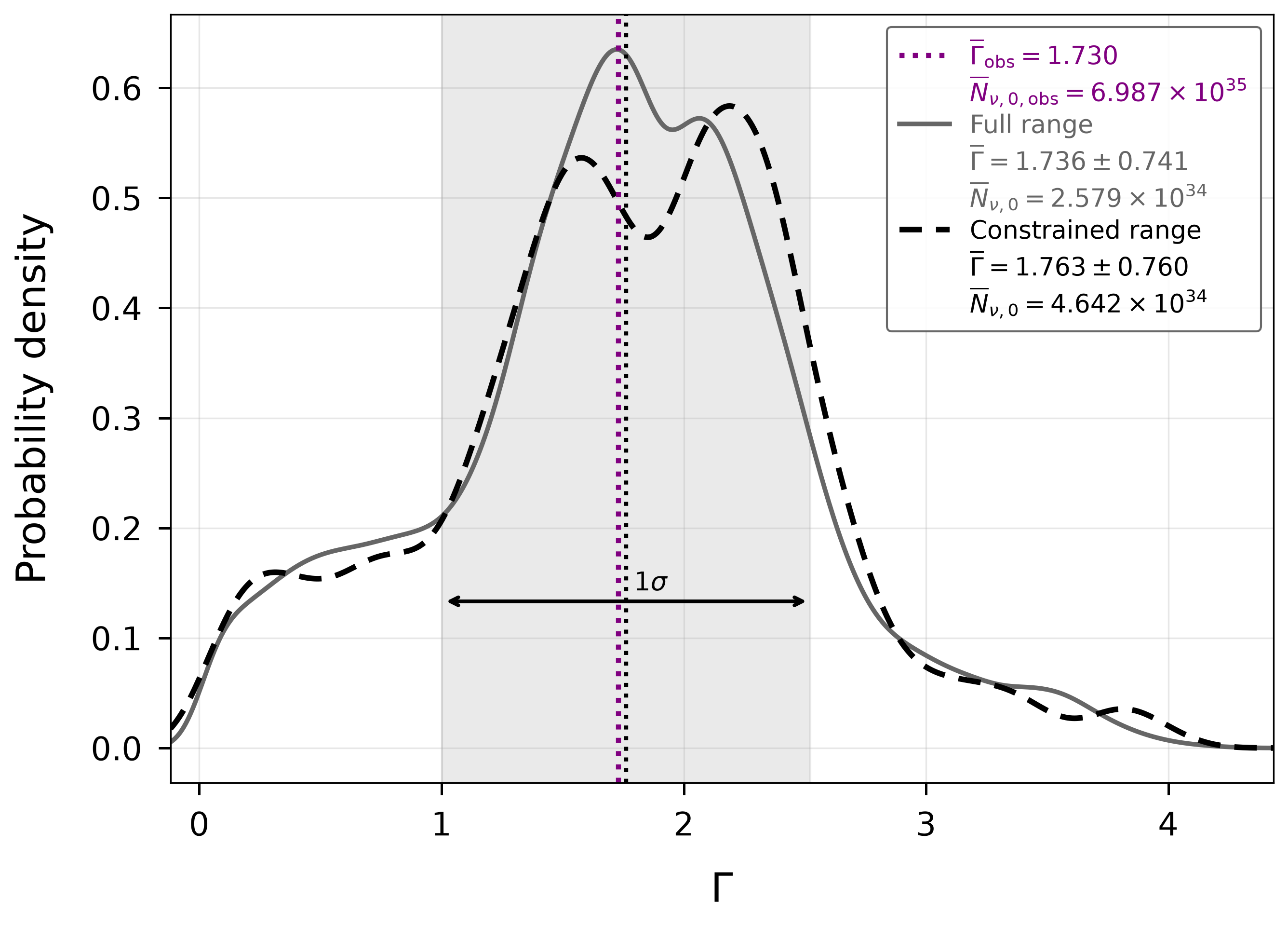}
    \caption{The probability distribution of the photon index $\Gamma$ for the combined solutions from all three AGNs. The solid gray curve shows the distribution for all physical solutions corresponding to the full parameter ranges in Table \ref{tab:parameter_ranges}, while the dashed black curve shows the distribution for constrained parameter ranges. The vertical dotted lines delineate the average of the best-fit observed photon indices $\Gamma_{\rm obs}$ in purple and average of the model prediction for the three AGN $\overline{\Gamma}$ for the constrained parameter ranges. The shaded region represents the interval $\overline{\Gamma}\pm\sigma$ for constrained parameter space. The legend summarizes the values for ${\overline\Gamma}_{\rm obs}$, ${\overline N}_{\nu,0,\rm {obs}}$, and the mean values of $\Gamma$ and $N_{\nu,0}$ before and after applying the parameter range constraints for each AGN. All $N_{\nu,0}$ values are given in $\rm photons\, s^
{-1}\, Hz^{-1}$.}
    \label{fig:gamma distribution all agns}
\end{figure}
Figure \ref{fig:Gamma distribution peragn} shows the $\Gamma$ distribution for each AGN separately, with a different colour for each of the three AGN (ARK120 - blue, MCG63015 - orange, and NCG4593 - green). The solid and dashed lines show the distributions for full and constrained $\alpha$, $\alpha\beta$ parameter range. Even without constraining the parameter range, the predicted values of $\overline{\Gamma}$ are in reasonable agreement with $\Gamma_{\rm obs}$. But after constraining the parameter range, the mean values even more closely match the best-fit $\Gamma_{\rm obs}$ for each AGN. For ARK120, $\overline{\Gamma}$ changes from $1.682$ to $1.728$
after applying the constraints on the $\alpha$ and $\alpha\beta$ parameter range,  compared with $\Gamma_{\rm obs}=1.733$. We use $\nu_*=10^{17} \rm Hz$ as a normalization frequency to calculate $N_{\nu,0}$  the normalization of the power law defined in equation (\ref{eq:N_nu}). The mean photon spectrum normalization $\overline{N}_{\nu,0}$ increases from $2.279\times10^{35} \,\rm photons\, s^
{-1}\, Hz^{-1}$ to $3.447\times10^{35}\,\rm photons\, s^
{-1}\, Hz^{-1}$, compared with $N_{\nu,0,\rm obs}=5.822\times10^{35}\,\rm photons\, s^
{-1}\, Hz^{-1}$.  For MCG63015, $\overline{\Gamma}$ changes from $1.726$ to $1.864$ and $\overline{N}_{\nu,0}$ increases from $1.784\times10^{34}\,\rm photons\, s^
{-1}\, Hz^{-1}$ to $4.135\times10^{34}\,\rm photons\, s^
{-1}\, Hz^{-1}$ after applying the constraints, compared with  $\Gamma_{\rm obs}=1.819$ and $N_{\nu,0,\rm obs}=4.423\times10^{34}\,\rm photons\, s^
{-1}\, Hz^{-1}$. For NCG4593 after constraints, $\overline{\Gamma}$ changes from $1.801$ to $1.673$, while $\overline{N}_{\nu,0}$ increases from $9.094\times10^{33}\,\rm photons\, s^
{-1}\, Hz^{-1}$ to $1.070\times10^{34}\,\rm photons\, s^
{-1}\, Hz^{-1}$,  compared with  $\Gamma_{\rm obs}=1.623$ and $N_{\nu,0,\rm obs}=2.090\times10^{34}\,\rm photons\, s^
{-1}\, Hz^{-1}$. This suggests that not only is the basic  physical picture---that  disc emission comes from radial  transport and internal dissipation while coronal emission comes from vertical buoyancy transported magnetic fields and energy dissipation above the disc---consistent with the observations, but introducing the additional physically motivated constraints on  $\alpha \beta$ and $\alpha$ improves the agreement further. 

Figure \ref{fig:gamma distribution all agns} shows the same comparison after combining all the solutions from all three AGN, and treating them  as part of a statistical ensemble. Here, the predicted mean spectral index changes from $\overline{\Gamma}=1.736$ over the full parameter range to $\overline{\Gamma}=1.763$ for the constrained parameter range, as shown by the black dotted line. This can be compared to $\overline{\Gamma}_{\rm obs}=1.73$, the average of the best-fit $\Gamma_{\rm obs}$ values from the three AGNs, shown by the purple dotted line. The corresponding mean normalization changes from $\overline{N}_{\nu,0}=2.579\times10^{34}\,\rm photons\, s^
{-1}\, Hz^{-1}$ to $\overline{N}_{\nu,0}=4.642\times10^{34}\,\rm photons\, s^
{-1}\, Hz^{-1}$, compared with $\overline{N}_{\nu,0,\rm obs}=6.987\times10^{35}\,\rm photons\, s^
{-1}\, Hz^{-1}$. The predicted $\overline{\Gamma}$ and $\overline{N}_{\nu,0}$ for individual AGN and combined solutions, for the full and constrained parameter spaces, are summarized in Table \ref{tab:parameter_ranges}.

\subsection{Application to a spectrum with one x-ray data point} \label{sec:single data point}
\begin{figure*}
    \centering

    \subfloat[]{
        \includegraphics[width=0.98\textwidth]
        {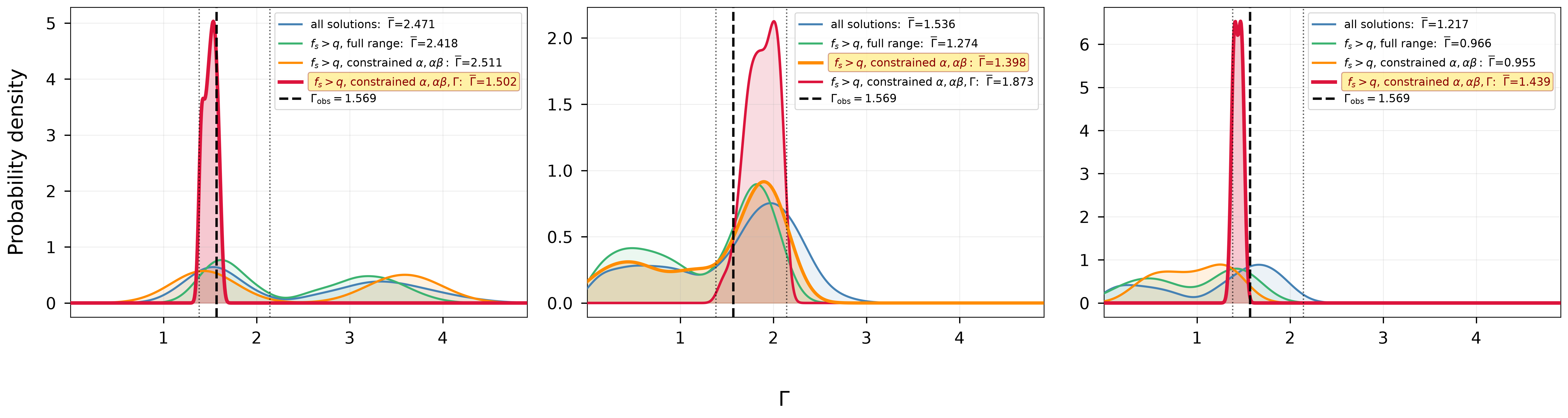}
        \label{fig:gamma_bins}
    }

    \vspace{0.3cm}

    \subfloat[]{
        \includegraphics[width=0.98\textwidth]
        {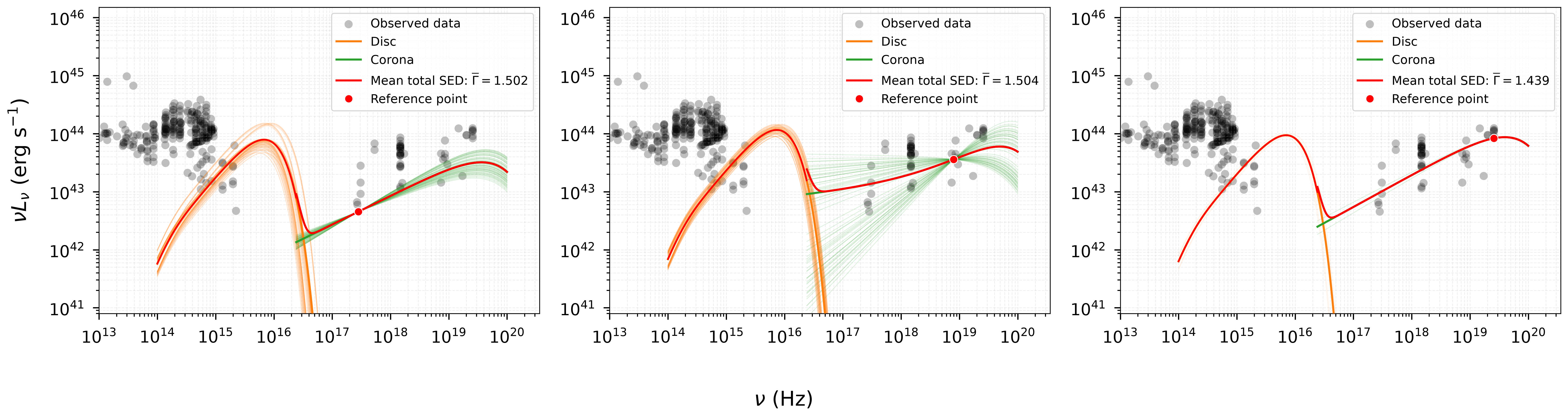}
        \label{fig:NGC5548sed}
    }
\caption{
Application of the model to NGC~5548 assuming that only one x-ray data reference point is available. (a) Distribution of predicted photon index $\Gamma$ resulting if the on data point is in the low-, middle-, and high-frequency regions of the x-ray spectra, respectively from left to right.  Curves show the results for all solutions (blue), physical solutions within our full parameter range (green), physical solutions within constrained parameter space (orange) and the latter solutions with the additional restriction from Table \ref{tab:parameter_ranges} $1.38\leq\Gamma\leq2.14$ (red). This restricted range is also indicated by the black dotted vertical lines.  The black dashed line marks $\Gamma_{\rm obs}$ for NGC~5548. The case with mean photon index closest to $\Gamma_{\rm obs}$ is highlighted in each legend. (b) Predicted SEDs  for each of the  cases corresponding to the chosen x-ray reference points, shown in red,  being from  one of the three distinct  frequency regions using the constrains highlighted in panel (a). Orange and green curves show the disc and coronal components, while the mean total SED is displayed in red. Gray points show the observed data points for comparison.
}
    \label{fig:NGC5548}
\end{figure*}
For the best results, the way we apply the model depends on whether we are studying a population of AGNs or an individual source and how much information is available from observations on their x-ray spectra.

For a study of a large sample of Seyfert 1 AGNs, we can assume that sources follow the same physics and therefore follow a similar x-ray power law. In this case, the $\overline{\Gamma}$ obtained from the three AGNs studied above can be used to predict x-ray spectra. It is important to  keep in mind that a much larger sample of AGNs with well-measured x-ray spectra would provide a better constrained $\Gamma$ range and $\overline{\Gamma}$ than the three AGNs considered here. A further improvement is also needed to reduce  variability and instrumental differences in x-ray data by separating observations  by the AGN activity epoch and telescope. This can lead to a more tightly constrained power-law and $\overline{\Gamma}$.

When modelling an individual AGN  with limited x-ray observations, our model can be used to predict the coronal power-law spectrum even when only one x-ray data point is available.
Here, we demonstrate this using Seyfert 1 AGN -  NGC5548. Even though  a well-sampled observed x-ray spectrum is available for NGC5548, we treat the source as if only one x-ray data point were observed and use that point in our model to predict the coronal power law. We repeat this using a few points selected from the different parts of the x-ray spectrum to test how much the position of the measurement influences a predicted power-law. 
The solutions are sensitive to the characteristic parameters of the AGN and in some cases we might get no solutions after applying the physical constrains. So, in this single reference point study, we consider two distinct observationally viable values for the black-hole mass of this object $M_{\rm BH}=4.0\times10^{7}M_\odot $ and $M_{\rm BH}=6.5\times10^{7}M_\odot$ and two values for the bolometric luminosity $L_{\rm bol}=2.6\times10^{44} \rm erg \ s^{-1} $ and $ L_{\rm bol}= 5.0\times10^{44} \rm erg \ s^{-1}$ \citep{Pancoast2014NGC5548,Pounds2003NGC5548,Bentz2007NGC5548,  Mehdipour2024NGC5548}. We use all four mass-luminosity combinations to increase a change of obtaining surviving solutions for a chosen data point.
The results shown in Figure \ref{fig:NGC5548} combines the solutions from all four mass-luminosity combinations. 

In panel (a) of Figure \ref{fig:NGC5548}, we group our chosen reference points into three frequency bins:  low-frequency region,
$7.4\times10^{16}\leq\nu\leq4.0\times10^{17}\,\mathrm{Hz}$;  middle-frequency region,
$4.0\times10^{17}<\nu\leq1.9\times10^{19}\,\mathrm{Hz}$ and  high-frequency region,
$1.9\times10^{19}<\nu\leq1.0\times10^{20}\,\mathrm{Hz}$ with one, four, and two reference points, respectively and look at how the resulting  $\Gamma$ distributions change as we apply physical $f_{\rm s}>q$ and $\alpha $, $ \alpha\beta$ parameter range constraints. We then compare the resulting distributions with the photon index $\Gamma_{\rm obs}$ obtained from the best-fitting power law to the full observed x-ray spectra of  NGC~5548. For the middle-frequency reference points, we see that just using our model with a narrower $\alpha,\alpha\beta$ parameter range,  already produces a close $\overline{\Gamma}$ to the best-fit value $\Gamma_{\rm obs}$. But for the low- and high-frequency points, the $\Gamma$ distributions are shifted to higher and lower values, respectively and we need to apply additional constraints on the allowed $\Gamma$ range obtained from the three AGNs studied above. Based on this behavior, we apply our model differently depending on which frequency range the  observed data point fits. For points near the low and high ends of the x-ray corona spectrum, we apply an additional constraint from the population of all sources that  $1.38\leq\Gamma\leq 2.14$, while for the middle-frequency range points we do not apply this constraint. Following this procedure, 
our model can be applied to predict the coronal power law from a single x-ray data point.  We offer this procedure as a generic one for any source with a single data point, with the key aspect of the procedure being distinguishing the distinct procedures depending on whether the data point comes from the middle frequency regime compared to the low and high reigmes. 
Panel (b) of Figure \ref{fig:NGC5548} shows predicted SEDs using the appropriate constraints for the two reference points near the edges of the x-ray spectra and one from the middle. The three SED panels, from left to right, use reference-points with
$(\nu,\nu L_\nu)=(2.827260\times10^{17},\,4.527061\times10^{42})$,
$(7.993620\times10^{18},\,3.596450\times10^{43})$, and
$(2.573010\times10^{19},\,8.303906\times10^{43})$,
respectively, in units of
$(\mathrm{Hz},\,\mathrm{erg\,s^{-1}})$. For each point, the surviving disc and coronal components are shown in orange and green colors, while the red curve represents the mean total SED.

\subsection{Connection of power law index $\Gamma$ to microphysics} \label{sec:microphyscis}

In the present paper we do not address the microphysical details of particle acceleration and radiative processes.  We have
assumed that the coronal spectrum is a power law, but our work here shows that the magnetic buoyancy driven corona formation paradigm is consistent with a power law that matches observations. To close the  loop from first principles, microphysical models of particle acceleration and radiation should independently produce power laws in the same range.  

For coronal plasma conditions,  microphysical calculations must follow how electrons gain energy through shocks, magnetic reconnection, or plasma turbulence, how they lose energy through radiative cooling \citep{Bell1978, Blandford+Eichler1987, Blackman+Field1994}.  Field rogers 1993 considered electron acceleration through first-order fermi acceleration at strong shocks in magnetic loops above the disc. Including synchrotron and inverse-Compton energy looses they obtained a power-law radiation spectrum  corresponding to a photon index $\Gamma=2$. \cite{Haardt+Maraschi1991} obtained similar slope in their two-phase model of a hot corona and cold accretion disc. In their model, inverse-Compton cooling of the corona is controlled by disc photons including the ones emitted when coronal x-rays heat up the disc. This feedback between disc and corona produces an x-ray power law with a photon index $\Gamma=2$.

More recently, 2D and 3D particle-in-cell (PIC) simulations have modeled relativistic magnetic reconnection in both electron-positron and electron-ion plasmas \citep{Sironi+Beloborodov2020,Sridhar2021,Sridhar2023}. Combining these PIC stimulation with Monte Carlo radiative transfer  \cite{Sridhar2021,Sridhar2023} obtained  $\Gamma\sim1.5$ from inverse-Compton scattering for both plasma compositions. \cite{Grovselj2024,Grovselj2026} instead used 3d and 2D radiative PIC simulations of a turbulent  electron-positron and electron-ion plasma. Both simulations include self-consistent Compton scattering, injection and escape of photons, while electron-ion simulations also include the injection and diffusive escape of charged particles. These simulations produced photon indices $\Gamma\sim1.6$ and $\Gamma\sim1.77$. The former was consistent with the hard state of the black hole binary Cyg X-1, while later is compared with the observations of NGC 4151 since it is  one of the x-ray brightest AGNs \citep{Ulrich2000}.

This variety of calculations have been carried out under different assumptions about plasma composition and energy dissipation, but they do produce photon indices between $1.5\lesssim\Gamma\lesssim2$,   consistent with the photon index range that emerges from our  model and the observations. A more ambitious comparison could be eventually be made using global numerical simulations that invoke the coronal conditions assumed in our model and also study in detail the microphysical particle acceleration via subgrid modeling.

\section{Conclusion} \label{sec:conclusion}

In this work, we tested the plausibility of the conceptual paradigm that Seyfert AGN  coronae are powered from below only by magnetic fields whose scales are large enough to escape turbulent shredding on their buoyant rise time. We have combined mean field accretion disc theory, which includes both local radial and vertical angular momentum transport into a corona and outflows, with constraints from magnetic buoyancy  to connect vertically transported magnetic energy to the coronal emission and spectra. Extracting magnetic energy spectra from previous MHD simulations, we have quantified the fraction of magnetic energy contained in large scale structures that survive the buoyant rise and relate it to the fraction of coronal to total luminosity. Together with the assumed thermal disc radiation and a power law  coronal radiation spectrum, this allows us to predict the coronal photon index, without explicitly calculating particle acceleration and radiative processes.

Using available observations and x-ray data for the Seyfert AGNs ARK120, MCG63015 and NCG4593, we find that physically motivated restrictions applied to the model parameters improve the agreement between the predicted mean and observationally derived photon indices for each source, supporting our theoretical model framework. The resulting predicted mean photon spectral index is $\overline{\Gamma}=1.76\pm0.76$, also in agreement with the mean observed value of $\overline{\Gamma}_{\rm obs}=1.73$. This, in turn,  gives utility to the combined mean photon index, if we assume Seyfert AGNs are similar members of a statistical ensemble.  We also use NGC5548 to show how the model can be applied to the AGN even with a single observed x-ray data point to predict the coronal power law.

Extending this study and approach for a larger sample of AGNs with well measured x-ray data will help to constrain the mean photon index. Our results also presently have a large standard deviation which can be reduced in further work by separating observations by AGN activity epoch and telescope to account for the variability and instrumental differences. 

The overall agreement of our results with observations supports the specific large scale magnetic buoyancy paradigm for Seyfert AGN corona formation. Further validation requires  full first-principles  microphysical calculations of particle acceleration and radiation which independently produce power laws in the same range as those predicted in our framework and found in observations.
\section*{Data Availability}

All the data used in the paper is available at the NASA/IPAC Extragalactic Database (NED) https://ned.ipac.
caltech.edu or is given in Table \ref{tab:agn}.

\appendix 
\counterwithin{figure}{section}
\renewcommand{\thefigure}{\thesection\arabic{figure}}
\renewcommand{\theHfigure}{appendix.\thesection.\arabic{figure}}

\appendix \label{app:p}
\section{The radial mass-loss index $p$ }\label{app:p} 
The radial mass-loss index $p$ is determined by requiring that the total accretion power predicted by the outflow-modified accretion-disc model reproduces the observed bolometric luminosity of each AGN. The total accretion power is given by Equation~(\ref{eq:power total}).

The outer mass accretion rate is estimated from the observed bolometric luminosity and estimating radiative efficiency $\eta$ form the spin $a$.

\begin{equation}
\dot{M}_{\rm out}
=
\frac{L_{\rm bol}}{\eta c^2},
\label{eq:mdot_lbol}
\end{equation}

where $L_{\rm bol}$ is the observed bolometric luminosity.

Substituting Equation~(\ref{eq:mdot_lbol}) into Equation~(\ref{eq:power total}) gives

\begin{equation}
P_a(p)
=
\frac{GM}{2r_i}
\frac{L_{\rm bol}}{\eta c^2}
\left(\frac{r_{\rm out}}{r_i}\right)^{-p}
\left(\frac{1+2p}{1-p}\right).
\label{eq:pa_sub}
\end{equation}

The wind parameter is obtained by requiring that the model accretion power equals the observed bolometric luminosity,

\begin{equation}
P_a(p)=L_{\rm bol}.
\label{eq:condition}
\end{equation}

Substituting Equation~(\ref{eq:pa_sub}) into Equation~(\ref{eq:condition}) yields

\begin{equation}
L_{\rm bol}
=
\frac{GM}{2r_i}
\frac{L_{\rm bol}}{\eta c^2}
\left(\frac{r_{\rm out}}{r_i}\right)^{-p}
\left(\frac{1+2p}{1-p}\right).
\end{equation}

Since $L_{\rm bol}\neq 0$, it cancels from both sides, giving

\begin{equation}
\frac{GM}{2r_i\eta c^2}
\left(\frac{r_{\rm out}}{r_i}\right)^{-p}
\left(\frac{1+2p}{1-p}\right)
=
1.
\label{eq:p_equation}
\end{equation}

For convenience, we define the dimensionless ratio

\begin{equation}
F_{\rm model}(p)
=
\frac{GM}{2r_i\eta c^2}
\left(\frac{r_{\rm out}}{r_i}\right)^{-p}
\left(\frac{1+2p}{1-p}\right),
\label{eq:fmodel}
\end{equation}

such that the physical solution satisfies

\begin{equation}
F_{\rm model}(p)=1.
\end{equation}

\begin{figure}
    \centering
    \includegraphics[width=\columnwidth]{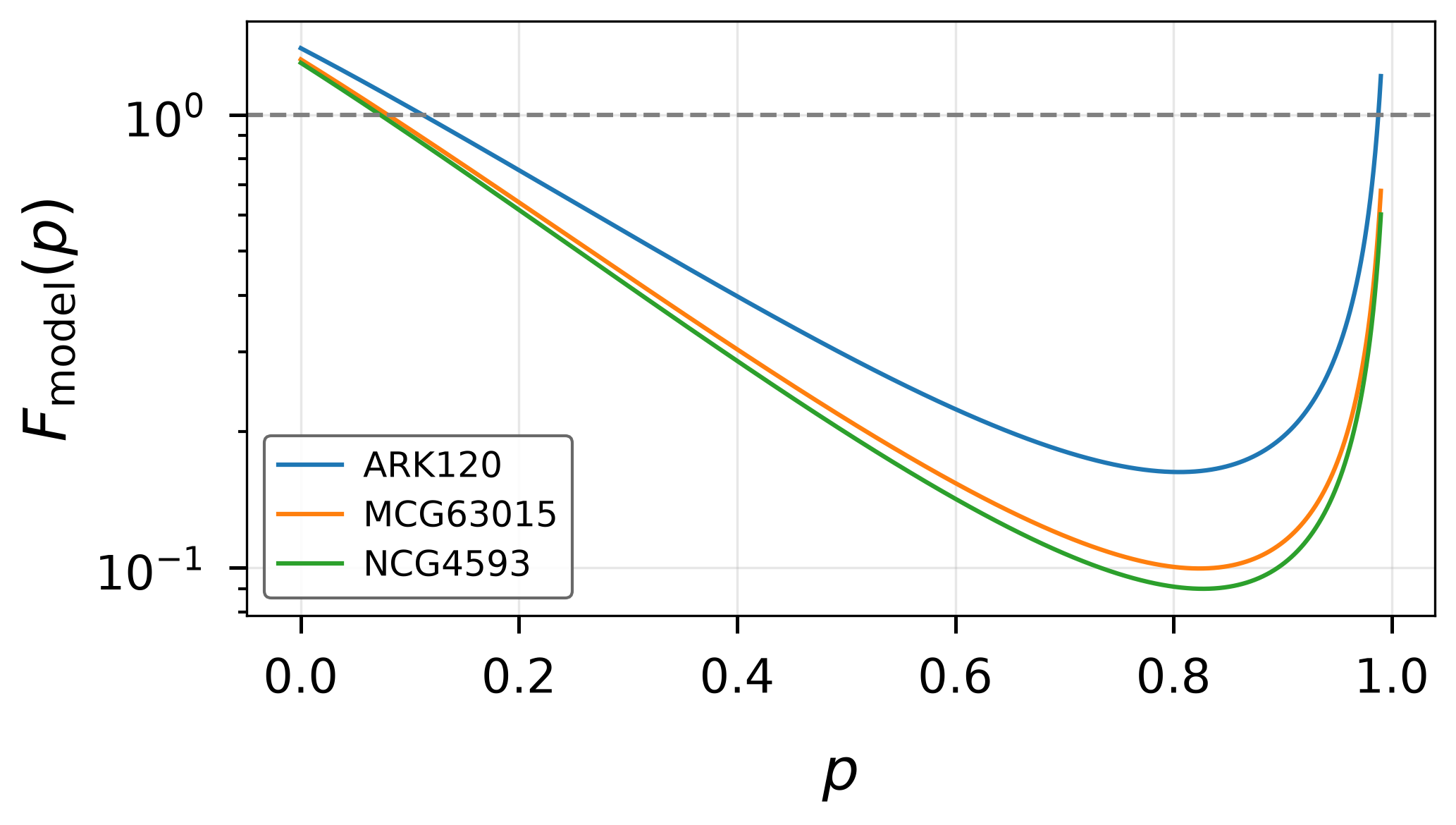}
    \caption{Solid curved show $F_{\rm model}(p)$ as a function of the radial mass loss index $p$ for ARK120 (blue), MCG63015 (orange) and NCG4593 (green). The dashed like marks $F_{\rm model}(p)=1$. Intersections between the solid curves and dashed line determines the value of $p$ for each AGN for which the model accretion power equals the observed bolometric luminosity of each AGN.}
    \label{fig:F_vs_p}
\end{figure}

Figure~\ref{fig:F_vs_p} shows $F_{\rm model}(p)$ as a function of $p$ for each AGN. The horizontal dashed line denotes $F_{\rm model}=1$, and the corresponding value of $p$ is obtained from the intersection of each curve with this line.

\section{The magnetic energy fraction $f_s$}\label{app:f_s} 
The fraction of disc magnetic energy $f_s$ with scales larger than $l_c$ is computed using the magnetic energy spectra from the numerical simulations of magnetized turbulent accretion discs studied by \cite{Held+Mamatsashvili+Pessah2024}. 

\begin{figure}
    \centering
    \includegraphics[width=\columnwidth]{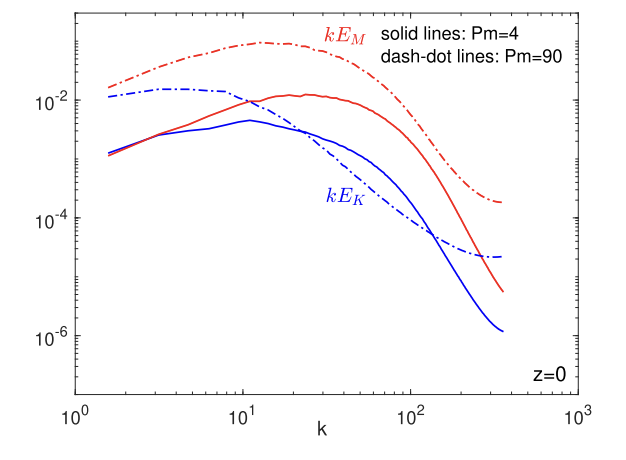}
    \caption{Ring-averaged magnetic ($kE_M$, red) and kinetic ($kE_K$, blue) energy spectra as a function of wavenumber $k$ at the disc midplane ($z=0$), taken from \cite{Held+Mamatsashvili+Pessah2024}. Solid and dash-dotted lines correspond to magnetic Prandtl numbers $\mathrm{Pm}=4$ and $\mathrm{Pm}=90$, respectively.}
    \label{fig:fig11}
\end{figure}

We use the ring-averaged magnetic energy spectra shown in Figure \ref{fig:fig11}, where the quantity  $kE(k)$ is first converted to the energy spectrum $E(k)=kE(k)/k$, before computing $f_s$ using equation (\ref{eq:fs}).

We consider two physically motivated critical wavenumbers corresponding to the minimum and maximum buoyant velocity limits.
For the maximum buoyant velocity limit considered in this work, $U_{b, \rm max}=c_s$, the critical length scale is $l_c=\alpha^{1/2} H$
while for the minimum buoyant velocity $U_{b, \rm min}=\alpha c_s$, the critical length scale is $l_c= H$, where $H$ is the disc scale height.
The corresponding critical wavenumbers are:
\begin{equation}
k_{\rm crit}=
\begin{cases}
2 \pi/\alpha^{1/2} H, & U_{b, \rm max}=c_s,\\[4pt]
2 \pi/H, & U_{b, \rm min}=\alpha c_s,
\end{cases}
\end{equation}
First, the total magnetic energy is calculated by integrating the spectrum over the full range of wavenumbers. Next the spectrum is integrated from the lowest wavenumber up to the critical wavenumber $k_{\rm crit}$. The fraction $f_s$ is calculated by taking the ratio of these two integrals.  This calculation is done for different sets of values of $\alpha$ and $\alpha \beta$ given in table \ref{tab:parameter_ranges}.
In this study we evaluate all integrals for magnetic Prandtl number $P_{\rm m}=4$ at mid-plane $z=0$.

\section{The Coronal Photon Index $\Gamma$}\label{app:gamma}
To model the coronal x-ray spectrum, we use a power law with an exponential high-energy cut-off given by equation (\ref{eq:N_nu}). The normalized coronal spectrum can be written as 
\begin{equation}
\nu L_{\nu} =
(\nu L_{\nu})_0
\left(\frac{\nu}{\nu_0}\right)^{2-\Gamma}
\exp\left[-\frac{\nu-\nu_0}{\nu_{\rm cut}}\right].
\label{eq:normalized_corona}
\end{equation} 
where $(\nu_0,(\nu L_{\nu})_0)$ is observed x-ray point in the rest frame. So the power law has to pass through the observed data point. The integrated luminosity under a power law  $I(\Gamma)$  has to match the coronal luminosity calculated from the equation (\ref{eq:L_c}), which constrains the allowed values for the photon index.
\begin{align}
I(\Gamma)
&=
\int_{\nu_{\min}}^{\nu_{\max}}
L_{\nu}\,d\nu  \nonumber \\
&= (\nu L_{\nu})_0
\nu_0^{\Gamma-2}
\exp\left(\frac{\nu_0}{\nu_{\rm cut}}\right)
\int_{\nu_{\min}}^{\nu_{\max}}
\nu^{1-\Gamma}
\exp\left(-\frac{\nu}{\nu_{\rm cut}}\right)d\nu .
\label{eq:I_gamma_integral}
\end{align}
where $\nu_{\min}$ and $\nu_{\max}$ represent minimum and maximum frequency values for the x-ray coronal band. In the study we have taken$\nu_{\min}=2.4\times10^{16}$ Hz and $\nu_{\max}=4.0\times10^{20}$ Hz for all AGNs.

This integral can be expressed in terms of the incomplete gamma function if we introduce  $t=\nu/\nu_{\rm cut}$, $s=2-\Gamma$,
$x_1=\nu_{\min}/\nu_{\rm cut}$, and
$x_2=\nu_{\max}/\nu_{\rm cut}$ it gives
\begin{equation}
I(\Gamma)
=
(\nu L_{\nu})_0
\left(\frac{\nu_{\rm cut}}{\nu_0}\right)^{2-\Gamma}
e^{\nu_0/\nu_{\rm cut}}
\mathcal{G}(2-\Gamma;x_1,x_2),
\label{eq:I_gamma_final}
\end{equation}
where
\begin{equation}
\mathcal{G}(s;x_1,x_2)
=
\int_{x_1}^{x_2}t^{s-1}e^{-t}\,dt .
\end{equation}
For $s>0$ ($\Gamma<2$), this finite-interval integral can be written as
$\gamma(s,x_2)-\gamma(s,x_1)$ in terms of lower incomplete gamma
functions, while for $s<0$ ($\Gamma>2$) it may be written as
$\Gamma_{\rm u}(s,x_1)-\Gamma_{\rm u}(s,x_2)$ using upper incomplete
gamma functions. At $s=0$ ($\Gamma=2$), the corresponding expression is
$E_1(x_1)-E_1(x_2)$.

\begin{equation}
\mathcal{G}(2-\Gamma;x_1,x_2)
=
\begin{cases}
\gamma(2-\Gamma,x_2)-\gamma(2-\Gamma,x_1),
& \Gamma<2,\\[5pt]
E_1(x_1)-E_1(x_2),
& \Gamma=2,\\[5pt]
\Gamma_{\rm u}(2-\Gamma,x_1)
-\Gamma_{\rm u}(2-\Gamma,x_2),
& \Gamma>2.
\end{cases}
\label{eq:app_piecewise_Gamma}
\end{equation}

\begin{figure}
    \centering
    \includegraphics[width=0.88\linewidth]
    {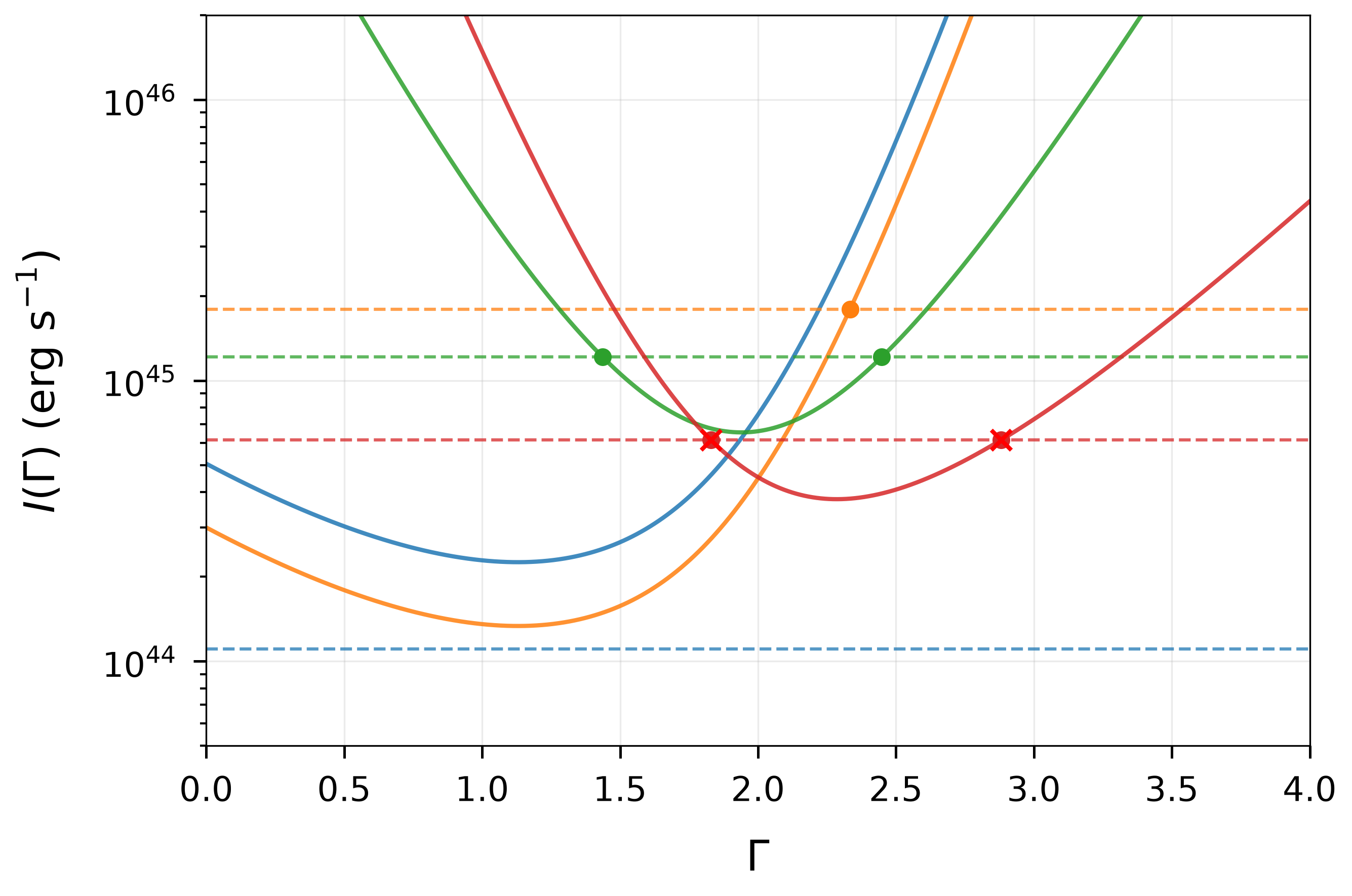}
    \caption{Solid curves represent integrated coronal luminosity $I(\Gamma)$ for selected ARK120 AGN cases, illustrating zero, one, or two solutions. The horizontal dashed lines indicate the corresponding coronal luminosities calculated from equation (\ref{eq:L_c}). Filled circles mark solutions found from $I(\Gamma)=L_c$; red crosses indicate solutions discarded by the physical constraint $f_s>q$. }
    \label{fig:ark120_gamma_roots}
\end{figure}

Figure \ref{fig:ark120_gamma_roots} displays the integrated coronal luminosity given by equation (\ref{eq:I_gamma_final}) as a function of the photon index $\Gamma$. The horizontal dashed lines indicate the coronal luminosities calculated from equation (\ref{eq:L_c}) for one specific case and the observed x-ray data point used for normalization. $\Gamma$ is determined form $I(\Gamma)=L_c$ condition. As we can see from Figure \ref{fig:ark120_gamma_roots}, depending on the case,  we can have no solution, one solution or two solutions for $\Gamma$. Note that some solutions satisfy the $I(\Gamma)=L_c$ condition, but later they are discarded and are therefore crossed with red lines in Figure \ref{fig:ark120_gamma_roots}, since they correspond to the non-physical cases where $f_s \leq q$.

\section*{Acknowledgments}
KK acknowledges support from a Horton Graduate Fellowship from the Laboratory for Laser Energetics.  We acknowledge support from National Science Foundation grant 
PHY-2020249, which funds the Center for Matter at Atomic Presure (CMAP). This material is based upon work supported by the Department of Energy [National Nuclear Security Administration] University of Rochester “National Inertial Confinement Fusion Program” under Award Number(s) DE-NA0004144.

\bibliographystyle{apsrev4-1}

\bibliography{agn}

\end{document}